\documentclass[useAMS,usenatbib]{mn2e}
\usepackage{longtable,lscape}
\usepackage{amsmath}

\usepackage{graphicx}
\usepackage{aas_macros}
\usepackage{amssymb}

\newcommand{\about}{$\sim\!\!$~}
\newcommand{\kms}{km~s$^{-1}$}
\newcommand{\etal}{et~al.\ }

\newcommand{\vneb}{$v_\textrm{neb}$}

\newcommand{\bvri}{\protect\hbox{$BV\!RI$} }

\newcommand{\bvmax}{\protect\hbox{$\left(B-V\right)_{\rm max}$}}

\mathchardef\mhyphen="2D

\newcommand{\be}{\begin{displaymath}}
\newcommand{\ee}{\end{displaymath}}

\def\lsim{\hbox{\rlap{\raise 0.425ex\hbox{$<$}}\lower 0.65ex\hbox{$\sim$}}}
\def\gsim{\hbox{\rlap{\raise 0.425ex\hbox{$>$}}\lower 0.65ex\hbox{$\sim$}}}

\newcommand{\ion}[2]{#1$\;${\small{#2}}\relax}

\graphicspath{{/Users/jsilv/BSNIP/late_time/plots/}}

\title[Late-Time SN~Ia Spectra]{Berkeley Supernova Ia Program V:
  Late-Time Spectra of Type Ia Supernovae} 

\author[Silverman, et~al.]{Jeffrey~M.~Silverman,$^{1,2}$\thanks{E-mail:
    jsilverman@astro.as.utexas.edu} Mohan~Ganeshalingam,$^{1,3}$ Alexei~V.~Filippenko$^{1}$ \\
$^{1}$Department of Astronomy, University of California, Berkeley, CA 94720-3411, USA \\
$^{2}$Department of Astronomy, University of Texas, Austin, TX 78712-0259, USA \\
$^{3}$Lawrence Berkeley National Laboratory, Berkeley, CA 94720, USA\\
}

\begin{document}
\date{Accepted  . Received   ; in original form  }
\pagerange{\pageref{firstpage}--\pageref{lastpage}} \pubyear{2012}
\maketitle
\label{firstpage}

\begin{abstract}
In this work we analyse late-time ($t > 100$~d) optical spectra of
low-redshift ($z < 0.1$) Type Ia supernovae (SNe~Ia) which come mostly
from the Berkeley Supernova Ia Program dataset. We also present
spectra of SN~2011by for the first time. The BSNIP sample studied
consists of 34 SNe~Ia with 60 nebular spectra, to which we add nebular
spectral feature measurements of 20 SNe~Ia from previously published
work (Maeda \etal 2011; Blondin \etal 2012), representing the
largest set of late-time SN~Ia spectra ever analysed. The full width
at half-maximum intensity (FWHM) and velocities of the [\ion{Fe}{III}]
$\lambda$4701, [\ion{Fe}{II}] $\lambda$7155, and [\ion{Ni}{II}]
$\lambda$7378 emission features are measured in most observations of
spectroscopically normal objects where the data have signal-to-noise
ratios $\ga 20$~px$^{-1}$ and are older than 160~d past maximum
brightness. The velocities of all three features are seen to be
relatively constant with time, increasing only a few to
\about20~\kms~d$^{-1}$. The nebular velocity (\vneb, calculated by
taking the average of the [\ion{Fe}{II}] $\lambda$7155 and
[\ion{Ni}{II}] $\lambda$7378 velocities) is correlated with
the near-maximum-brightness velocity gradient and early-time ejecta
velocity. Nearly all high velocity gradient objects have redshifted
nebular lines while most low velocity gradient objects have
blueshifted nebular lines. No correlation is found between \vneb\ and
$\Delta m_{15}(B)$, and for a given light-curve shape there is a large
range of observed nebular velocities. The data also indicate a
correlation between observed \bvmax\ and \vneb.
\end{abstract}



\section{Introduction}\label{s:intro}

Type~Ia supernovae (SNe~Ia) can be used to accurately measure
cosmological parameters \citep[e.g.,][]{Astier06, Riess07,
  Wood-Vasey07, Hicken09:cosmo,
  Kessler09,Amanullah10,Conley11,Suzuki12}, and they led to the discovery
of the accelerating expansion of the Universe
\citep{Riess98:lambda,Perlmutter99}. SNe~Ia are thought to be the
result of thermonuclear explosions of C/O white dwarfs (WDs) (e.g., 
\citealt{Hoyle60, Colgate69, Nomoto84}; see \citealt{Hillebrandt00}
for a review). Despite their cosmological utility, we are still
missing a detailed understanding of the progenitor systems and
explosion mechanisms \citep[see][for further information]{Howell11}. 

By about 100~d past maximum brightness, SN~Ia ejecta have expanded
significantly and the SN enters the so-called ``nebular phase'' (as
opposed to the ``early-time'' phase). At these late
epochs, SN~Ia spectra consist of broad emission lines of (mostly)
iron-group elements (IGEs) and can yield valuable insights into the
physics of the explosion itself. However, SNe~Ia often appear quite
faint at these late phases and thus not many late-time spectra exist
in the literature. 

There are a handful of relatively ``normal'' SNe~Ia \citep[i.e., ones
that follow the light-curve width versus luminosity
relationship;][]{Phillips93} that have published spectra at phases
later than \about100~d \citep[e.g.,][]{Stritzinger07,Stanishev07,Leloudas09}. In
addition, there are some rare and peculiar SNe~Ia with published
late-time spectra 
\citep[e.g.,][]{Jha06:02cx,Silverman11}. While there are large
spectroscopic samples of SNe~Ia published
\citep{Matheson08,Blondin12,Silverman12:BSNIPI}, only a tiny fraction
of those data consist of late-time spectra. However, moderate-sized
comparative studies of \about14--24 SNe~Ia with late-time spectra have
been undertaken \citep{Mazzali98,Maeda10,Blondin12}.

\begin{figure}
\centering
\includegraphics[width=3.4in]{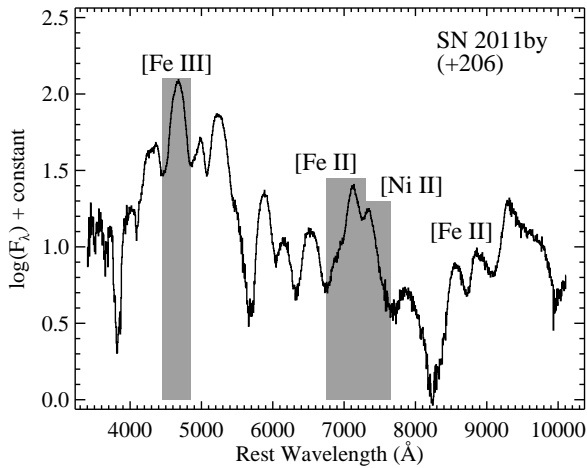}
\caption{Spectrum of SN~2011by from 206~d past maximum brightness with
prominent features labeled. The left-hand shaded area represents the
[\ion{Fe}{III}] $\lambda$4701 feature while the right-hand shaded area
represents the [\ion{Fe}{II}] $\lambda$7155 and [\ion{Ni}{II}]
$\lambda$7378 features.}\label{f:one_spec}
\end{figure}

These works have concentrated mainly on three broad emission features
centred near 4701, 7155, and 7378\,\AA\ (see
Figure~\ref{f:one_spec}). The
4701\,\AA\ feature  is likely a blend of various [\ion{Fe}{III}] lines
\citep{Stritzinger06,Maeda10c} which come from material produced by a
supersonic burning front \citep[i.e., a detonation;][]{Maeda10}. On
the other hand, the 7155\,\AA\ feature is likely due to
[\ion{Fe}{II}] \citep{Maeda10} and the 7378\,\AA\ feature is
probably from [\ion{Ni}{II}] \citep{Maeda10c}. These are both thought
to trace material from where the explosion began and where the burning
initially proceeded subsonically \citep[i.e., a
deflagration;][]{Maeda10}.

To the previous late-time SN~Ia spectral studies we now add this work, 
where we analyse 60 nebular-phase ($t >100$~d) low-resolution optical 
spectra of 34 low-redshift ($z < 0.1$) SNe~Ia obtained as part of the
Berkeley SN~Ia Program \citep[BSNIP;][]{Silverman12:BSNIPI}. The data
are presented by \citet{Silverman12:BSNIPI} (supplemented herein with 
one new object: SN~2011by),\footnote{All of the BSNIP data, along with
the SN~2011by data, will be made public on 1~Jan.~2013 via the
SuperNova DataBase \citep[SNDB;][]{Silverman12:BSNIPI}.}
and we utilise spectral feature measurement tools 
similar to those described by \citet{Silverman12:BSNIPII}. The
spectral data used in this work are summarised in
Section~\ref{s:data}, 
and our method of spectral feature measurement and our determination
of nebular velocities are described in
Section~\ref{s:procedure}. Comparison of our late-time spectral
measurements to other SN~Ia observables can be found in
Section~\ref{s:analysis}. We present our conclusions in
Section~\ref{s:conclusions}.


\section{Dataset}\label{s:data}

The late-time spectral sample used in this work is a subset of spectra
presented by \citet{Silverman12:BSNIPI}, with the addition of one new
object (see Section~\ref{ss:sn2011by}). The majority of the spectra were
obtained using the Shane 3~m telescope at Lick Observatory with the
Kast double spectrograph \citep{Miller93} and the 10~m Keck telescopes 
with the Low Resolution Imaging Spectrometer
\citep[LRIS;][]{Oke95}. The Kast data typically cover
3300--10,400~\AA\ with resolutions of \about11 and \about6~\AA\ on the
red and blue sides (crossover wavelength \about5500~\AA),
respectively. The LRIS spectra usually have a range of 3100--9200~\AA\
with resolutions of \about7 and \about5.5~\AA\ on the red and blue
sides (crossover wavelength \about5600~\AA), respectively.

All data were reduced using standard reduction methods. For more
information regarding the observations and data reduction, see
\citet{Silverman12:BSNIPI}. The spectral ages of the BSNIP data
referred to throughout this work are calculated using the redshift and
Julian Date of maximum presented in Table~1 of
\citet{Silverman12:BSNIPI}. Furthermore, photometric parameters 
(such as light-curve width and colour information) used in the present
study can be found in \citet{Ganeshalingam10:phot_paper}.

The BSNIP sample contains 81 spectra of 43 SNe~Ia older than 100~d
past maximum brightness. In order to measure the three emission
features mentioned above (i.e., [\ion{Fe}{III}] $\lambda$4701, 
[\ion{Fe}{II}] $\lambda$7155, and [\ion{Ni}{II}] $\lambda$7378),
we required that the spectra span
\about4300--7250~\AA, which decreased the sample to 73 spectra of 38
SNe~Ia. In previous BSNIP studies \citep{Silverman12:BSNIPII}, we {\it
  a priori} ignored the extremely peculiar SN~2000cx
\citep[e.g.,][]{Li01:00cx}, SN~2002cx
\citep[e.g.,][]{Li03:02cx,Jha06:02cx}, SN~2005gj
\citep[e.g.,][]{Aldering06,Prieto07}, and SN~2005hk
\citep[e.g.,][]{Chornock06,Phillips07}. These objects are so
spectroscopically peculiar that it is difficult to consistently
measure their spectral features in comparison to the bulk of the SN~Ia
population. Thus, in this work we will only concentrate on the SNe~Ia
that follow the ``Phillips relation'' \citep{Phillips93} and can
be used as cosmological distance indicators. We also include the
underluminous SN~1991bg-like objects, which lie on a quadratic
relationship between light-curve width and luminosity
\citep[e.g.,][]{Filippenko92:91bg,Leibundgut93}.

After removing the peculiar objects mentioned above, we are left with
33 SNe~Ia with 58 BSNIP spectra having $t > 100$~d. Of these, 7 were
in the late-time sample studied by \citet{Maeda10}, and those 7 plus 2
more were in the 
sample of \citet{Blondin12}. Each of the objects studied herein and the
spectral phases are listed in Table~\ref{t:v_neb}, along with
the full width at half-maximum intensity (FWHM) and velocity of the
[\ion{Fe}{III}] $\lambda$4701 feature, and the velocities of the
[\ion{Fe}{II}] $\lambda$7155 and [\ion{Ni}{II}] $\lambda$7378 features.

\setlength{\tabcolsep}{0.07in}
\onecolumn
\begin{landscape}
\begin{center}
\begin{longtable}{lcrrrr|lcrrrr}
\caption{\hbox to 6.1in{Nebular feature measurements. All FWHMs and velocities are in \kms. Uncertainties are given in parentheses.}}\label{t:v_neb} \\[-2ex]
\hline \hline
   &    & \multicolumn{1}{c}{FWHM} & \multicolumn{1}{c}{Velocity} & \multicolumn{1}{c}{Velocity} & Velocity\hspace{.13in} & & & \multicolumn{1}{c}{FWHM} & \multicolumn{1}{c}{Velocity} & \multicolumn{1}{c}{Velocity} & \multicolumn{1}{c}{Velocity} \\
SN Name & Phase$^\textrm{a}$ & \multicolumn{1}{c}{[Fe III] $\lambda$4701} & \multicolumn{1}{c}{[Fe III] $\lambda$4701} & \multicolumn{1}{c}{[Fe II] $\lambda$7155} & [Ni II] $\lambda$7378 & SN Name & \multicolumn{1}{c}{Phase$^\textrm{a}$} & \multicolumn{1}{c}{[Fe III] $\lambda$4701} & \multicolumn{1}{c}{[Fe III] $\lambda$4701} & \multicolumn{1}{c}{[Fe II] $\lambda$7155} & \multicolumn{1}{c}{[Ni II] $\lambda$7378} \\
\hline
\endfirsthead
\multicolumn{12}{c}{{\tablename} \thetable{} --- Continued} \\
\hline \hline
   &    & \multicolumn{1}{c}{FWHM} & \multicolumn{1}{c}{Velocity} & \multicolumn{1}{c}{Velocity} & Velocity\hspace{.13in} & & & \multicolumn{1}{c}{FWHM} & \multicolumn{1}{c}{Velocity} & \multicolumn{1}{c}{Velocity} & \multicolumn{1}{c}{Velocity} \\
SN Name & Phase$^\textrm{a}$ & \multicolumn{1}{c}{[Fe III] $\lambda$4701} & \multicolumn{1}{c}{[Fe III] $\lambda$4701} & \multicolumn{1}{c}{[Fe II] $\lambda$7155} & [Ni II] $\lambda$7378 & SN Name & \multicolumn{1}{c}{Phase$^\textrm{a}$} & \multicolumn{1}{c}{[Fe III] $\lambda$4701} & \multicolumn{1}{c}{[Fe III] $\lambda$4701} & \multicolumn{1}{c}{[Fe II] $\lambda$7155} & \multicolumn{1}{c}{[Ni II] $\lambda$7378} \\
\hline
\endhead

\hline \hline
\multicolumn{12}{l}{Continued on Next Page\ldots} \\
\endfoot

\hline \hline
\endlastfoot

SN 1989M               & 146.8 &     15000 (900) &   $-2800$ (160) &      $660$ (40) &      $450$ (30) &
SN 2001ep              & 114.9 &     12400 (900) &   $-3370$ (190) &        $\cdots$ &        $\cdots$ \\
SN 1991T               & 112.2 &    14300 (1500) &   $-2710$ (130) &        $\cdots$ &        $\cdots$ &
SN 2002bo              & 227.4 &    16600 (1300) &    $-1620$ (80) &    $2330$ (120) &    $2530$ (200) \\
SN 1991T               & 318.4 &    17200 (1700) &      $810$ (60) &        $\cdots$ &        $\cdots$ &
SN 2002cs              & 119.0 &    15500 (1300) &   $-3810$ (270) &        $\cdots$ &        $\cdots$ \\
SN 1991T               & 347.2 &    17000 (1400) &     $1810$ (90) &        $\cdots$ &        $\cdots$ &
SN 2002cs              & 143.5 &    15000 (1100) &   $-4280$ (220) &        $\cdots$ &        $\cdots$ \\
SN 1991bg              & 161.9 &      3200 (300) &   $-1840$ (110) &        $\cdots$ &        $\cdots$ &
SN 2002cs              & 172.2 &    15800 (1600) &   $-3720$ (220) &        $\cdots$ &        $\cdots$ \\
SN 1992G               & 110.8 &    14700 (1000) &   $-3000$ (200) &        $\cdots$ &        $\cdots$ &
SN 2002dp              & 103.1 &    14300 (1100) &   $-4080$ (250) &        $\cdots$ &        $\cdots$ \\
SN 1992G               & 111.9 &    14700 (1000) &   $-3130$ (260) &        $\cdots$ &        $\cdots$ &
SN 2002dp              & 134.7 &    13800 (1200) &   $-2630$ (120) &        $\cdots$ &        $\cdots$ \\
SN 1992G               & 126.8 &    16000 (1500) &   $-2990$ (160) &        $\cdots$ &        $\cdots$ &
SN 2002fk              & 148.7 &    14600 (1000) &   $-2650$ (160) &    $-1370$ (90) &   $-2750$ (200) \\
SN 1993Z               & 112.5 &    15100 (1100) &   $-2120$ (120) &        $\cdots$ &        $\cdots$ &
SN 2003gs              & 199.5 &    14700 (1200) &   $-2630$ (150) &    $3870$ (260) &     $1210$ (80) \\
SN 1993Z               & 132.4 &    15000 (1600) &   $-2120$ (130) &        $\cdots$ &        $\cdots$ &
SN 2004bv              & 135.2 &    14200 (1400) &   $-3300$ (280) &        $\cdots$ &        $\cdots$ \\
SN 1993Z               & 144.3 &    14800 (1200) &    $-1230$ (80) &        $\cdots$ &        $\cdots$ &
SN 2004bv              & 161.7 &    14300 (1100) &   $-2530$ (140) &        $\cdots$ &        $\cdots$ \\
SN 1993Z               & 161.2 &    14400 (1400) &   $-1350$ (100) &    $1820$ (100) &     $1180$ (90) &
SN 2004dt              & 109.2 &    13600 (1200) &   $-4970$ (250) &        $\cdots$ &        $\cdots$ \\
SN 1993Z               & 200.8 &    15300 (1100) &   $-2880$ (200) &    $1900$ (150) &    $3660$ (290) &
SN 2004dt              & 168.9 &        $\cdots$ &        $\cdots$ &   $-1470$ (100) &   $-4220$ (360) \\
SN 1993Z               & 232.7 &     14400 (900) &     $-720$ (50) &     $1000$ (60) &    $2390$ (190) &
SN 2005cf              & 317.1 &    15400 (1200) &     $-530$ (20) &      $410$ (40) &     $1400$ (80) \\
SN 1994D               & 114.7 &    15400 (1100) &   $-3010$ (170) &   $-2280$ (150) &   $-1150$ (100) &
SN 2005ke              & 362.1 &     10500 (700) &     $-430$ (30) &        $\cdots$ &        $\cdots$ \\
SN 1994ae              & 144.0 &    14500 (1200) &   $-2430$ (140) &        $\cdots$ &        $\cdots$ &
SN 2006D               & 126.5 &    14200 (1000) &   $-3320$ (200) &        $\cdots$ &        $\cdots$ \\
SN 1994ae              & 218.6 &     12200 (900) &     $-890$ (60) &        $\cdots$ &        $\cdots$ &
SN 2006X               & 126.3 &    15200 (1400) &   $-2970$ (160) &    $3420$ (230) &    $2320$ (160) \\
SN 1998bp              & 139.4 &      8700 (600) &   $-3490$ (220) &        $\cdots$ &        $\cdots$ &
SN 2006X               & 275.9 &    19300 (1400) &     $1210$ (70) &    $2610$ (170) &    $2730$ (220) \\
SN 1998bp              & 163.1 &     10600 (500) &   $-2340$ (150) &        $\cdots$ &        $\cdots$ &
SN 2006X               & 358.4 &    23700 (2500) &   $-3290$ (190) &    $1960$ (140) &    $3180$ (200) \\
SN 1998bu              & 235.1 &    15600 (1300) &    $-1030$ (50) &   $-1890$ (150) &     $-790$ (60) &
SN 2007af              & 113.8 &    13300 (1200) &   $-2920$ (140) &        $\cdots$ &        $\cdots$ \\
SN 1998bu              & 279.0 &    16400 (1100) &   $-2040$ (140) &   $-1710$ (150) &   $-1430$ (110) &
SN 2007af              & 121.7 &    13300 (1300) &   $-2790$ (190) &        $\cdots$ &        $\cdots$ \\
SN 1998bu              & 338.7 &    22200 (1900) &     $-650$ (40) &   $-2640$ (200) &   $-2330$ (220) &
SN 2007af              & 144.6 &    13100 (1100) &   $-3170$ (200) &        $\cdots$ &        $\cdots$ \\
SN 1998es              & 106.0 &    13500 (1300) &   $-3790$ (260) &        $\cdots$ &        $\cdots$ &
SN 2007af              & 158.5 &    13600 (1300) &   $-3550$ (210) &        $\cdots$ &        $\cdots$ \\
SN 1999aa              & 255.9 &    16700 (1500) &    $-1030$ (60) &      $830$ (40) &      $690$ (50) &
SN 2007fb              & 100.9 &    13700 (1300) &   $-3590$ (240) &        $\cdots$ &        $\cdots$ \\
SN 1999aa              & 281.5 &    18500 (1500) &      $520$ (30) &        $\cdots$ &        $\cdots$ &
SN 2007gi              & 153.3 &    15800 (1400) &    $-1440$ (90) &        $\cdots$ &        $\cdots$ \\
SN 1999ac              & 116.9 &     10300 (800) &   $-2200$ (120) &        $\cdots$ &        $\cdots$ &
SN 2007le              & 304.7 &    16700 (1600) &     $-230$ (10) &    $2160$ (150) &    $1890$ (160) \\
SN 1999by              & 183.5 &      9500 (700) &   $-1440$ (100) &        $\cdots$ &        $\cdots$ &
SN 2007sr              & 135.7 &    13700 (1200) &   $-2280$ (110) &        $\cdots$ &        $\cdots$ \\
SN 1999cw              & 104.6 &    14400 (1400) &   $-2920$ (170) &        $\cdots$ &        $\cdots$ &
SN 2008Q               & 199.3 &    15700 (1400) &    $-1860$ (90) &   $-2940$ (270) &   $-1450$ (120) \\
SN 1999cw              & 132.1 &     13600 (900) &   $-2410$ (150) &        $\cdots$ &        $\cdots$ &
SN 2011by              & 206.1 &    14500 (1300) &   $-2260$ (120) &   $-1350$ (100) &   $-1470$ (120) \\
SN 1999gh              & 118.0 &    12200 (1100) &   $-4200$ (260) &     $1290$ (90) &    $1840$ (130) &
SN 2011by              & 309.7 &    15300 (1200) &     $-280$ (20) &    $-1210$ (70) &   $-1750$ (150) \\
\hline \hline
\multicolumn{12}{p{9.2in}}{$^\textrm{a}$Phases of spectra are in rest-frame days relative to $B$-band maximum brightness using the heliocentric redshift and photometry reference presented in Table~1 of \citet{Silverman12:BSNIPI} and Section~\ref{ss:sn2011by} for SN~2011by.} \\
\end{longtable}
\end{center}
\end{landscape}
\twocolumn
\setlength{\tabcolsep}{6pt}

\subsection{SN~2011by}\label{ss:sn2011by}

To the BSNIP sample discussed above we also add new data for the
nearby Type Ia SN~2011by (see Appendix~\ref{s:appendix} for more
information). We obtained broad-band \bvri photometry as well as 7
near-maximum-brightness spectra and 2 late-time spectra of
SN~2011by. All of these data indicate that SN~2011by is
photometrically ($\Delta m_{15}(B) = 1.14 \pm 0.03$~mag) and
spectroscopically normal.

SN~2011by will serve as a high-quality individual case study that can
be directly compared to the larger late-time BSNIP sample discussed
above. Adding this object and its two late-time spectra to the
aforementioned dataset yields a sample of 60 spectra of 34 SNe~Ia with
$t > 100$~d past maximum brightness. This represents one of the
largest sets of late-time SN~Ia spectra ever analysed.

\section{Measuring Nebular Spectral Features}\label{s:procedure}

The algorithm used in this work to measure the [\ion{Fe}{III}]
$\lambda$4701, [\ion{Fe}{II}] $\lambda$7155, and [\ion{Ni}{II}]
$\lambda$7378 emission features is similar to that used to measure
absorption features in near-maximum-brightness BSNIP spectra. It is
described in detail by \citet{Silverman12:BSNIPII}, but here we give a
brief summary of the procedure. Each spectrum first has its
host-galaxy recession velocity removed and is corrected for Galactic
reddening \citep[according to the values presented in Table~1
of][]{Silverman12:BSNIPI}, and then is smoothed using a 
Savitzky-Golay smoothing filter \citep{Savitzky64}.

For each feature, a linear background is defined by connecting points
on either side of the feature and then subtracting out that background
flux level. A cubic spline is then fit to the smoothed data and
the expansion velocity is calculated from the wavelength at which the 
spline fit reaches its maximum. This maximum is also used to measure
the FWHM for the [\ion{Fe}{III}] $\lambda$4701 feature.

[\ion{Fe}{III}] $\lambda$4701 is usually the strongest feature in
late-time SN~Ia spectra and is often seen to be a single-peaked,
distinct feature. On the other hand, [\ion{Fe}{II}] $\lambda$7155 and
[\ion{Ni}{II}] $\lambda$7378 are found to be blended into either a
single- or double-peaked broad emission feature. Any spectrum where
the two features were severely blended with each other (or showed more
than two peaks near the wavelengths of interest) did not have
velocities measured for [\ion{Fe}{II}] $\lambda$7155 and
[\ion{Ni}{II}] $\lambda$7378. For the spectra that do show clearly
double-peaked profiles, a cubic spline is again fit to the smoothed
data and the expansion velocities are calculated from the wavelengths
at which the spline fit reaches its maximum on either side of the
local minimum (i.e., the point between the two peaks). The FWHM and
velocity of the [\ion{Fe}{III}] $\lambda$4701 feature and the
velocities of the [\ion{Fe}{II}] $\lambda$7155 and [\ion{Ni}{II}]
$\lambda$7378 features are displayed in Table~\ref{t:v_neb}. 

While we were able to measure [\ion{Fe}{III}] $\lambda$4701 in all
late-time spectra presented herein whose wavelength range included
this feature, we were only able to accurately measure the velocities
of [\ion{Fe}{II}] $\lambda$7155 and [\ion{Ni}{II}] $\lambda$7378 in 22
spectra of 15 SNe~Ia \citep[nebular velocities of 6 of these objects
have been calculated previously;][]{Maeda10,Blondin12}. In these
previous studies, velocities were reported even when there was only a
single peak near these wavelengths. With only one peak, however, it is
not clear which reference wavelength to use when calculating a nebular
velocity, and so for a velocity measurement in this work we require
that exactly two peaks be present in a late-time spectrum.

The underluminous SN~1991bg-like objects
\citep[e.g.,][]{Filippenko92:91bg,Leibundgut93} have narrower emission
profiles than do normal SNe~Ia at late times, but they do not typically show
the [\ion{Fe}{II}] $\lambda$7155 and 
[\ion{Ni}{II}] $\lambda$7378 features. Furthermore, the SN~1991T-like
objects \citep[e.g.,][]{Filippenko92:91T,Phillips92} and
SN~1999aa-like objects \citep[e.g.,][]{Li01:pec,Strolger02,Garavini04},
which together generally represent overluminous SNe~Ia, often have complex
and multi-peaked profiles near 7300~\AA, and thus we are again unable
to accurately measure the velocities of [\ion{Fe}{II}] $\lambda$7155
and [\ion{Ni}{II}] $\lambda$7378 for most of these objects. We are,
however, able to measure velocities in one spectrum of SN~1999aa. Note
that \citet{Maeda10} cut these peculiar objects from their sample as
well. Of the remaining ``normal'' SNe~Ia, 6 of the spectra in which we
are unable to reliably measure these features have low signal-to-noise
ratios (S/N $\la 20$~px$^{-1}$), with 3 of those having S/N~$<
5$~px$^{-1}$, and all normal SN~Ia spectra with S/N $> 20$~px$^{-1}$
were successfully measured.

\begin{figure}
\centering
\includegraphics[width=3.4in]{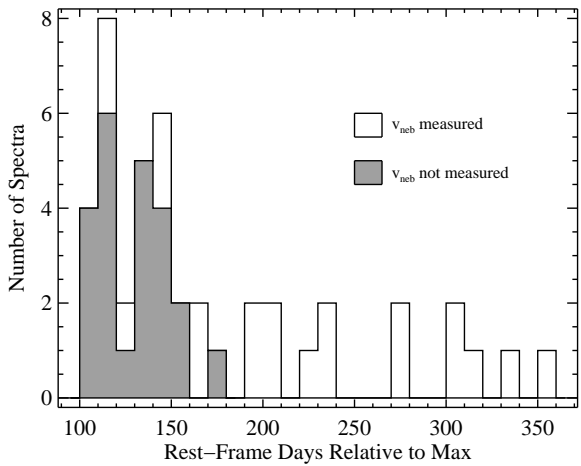} 
\caption{Histogram of spectral ages after removing peculiar objects
  and low-S/N spectra. The grey area represents spectra in which we
  are unable to measure [\ion{Fe}{II}] $\lambda$7155 and
  [\ion{Ni}{II}] $\lambda$7378 velocities. The white area represents
  spectra in which we can measure these velocities. Note that the vast
  majority of spectra with $t > 160$~d have measured
  velocities.}\label{f:age}
\end{figure}

Even after removing 9 spectra of peculiar SNe~Ia and 6 spectra with
low S/N, there are still 23 instances where we are unable to measure 
reliable velocities for the [\ion{Fe}{II}] $\lambda$7155 and
[\ion{Ni}{II}] $\lambda$7378 features. Nearly all of these failures are 
due to the spectra being obtained at too young a
phase. Figure~\ref{f:age} shows a histogram of the ages of the spectra
investigated in this work, {\it after} removing peculiar objects and
low-S/N spectra. The grey histogram represents the ages of spectra in
which we are unable to measure [\ion{Fe}{II}] $\lambda$7155 and
[\ion{Ni}{II}] $\lambda$7378 velocities, while the white histogram
represents ages where we successfully measure the velocities. At
epochs less than \about160~d past maximum brightness, the emission
near 7300~\AA\ is often still multi-peaked and too complex to
definitively identify the [\ion{Fe}{II}] $\lambda$7155 and
[\ion{Ni}{II}] $\lambda$7378 features. Thus, in order to be able to
measure these velocities, one should wait until \about160~d past
maximum, after which time we are able to measure the velocity in 16 of
17 spectra. Despite this finding, there {\it are} some BSNIP spectra at
115, 118, 126, 147, and 149~d past maximum that have measured
velocities.

The oldest spectrum in our sample is from 362~d past maximum
brightness, but it is SN~1991bg-like and we are unable to measure the
velocities of the [\ion{Fe}{II}] $\lambda$7155 and [\ion{Ni}{II}]
$\lambda$7378 features. The next-oldest spectrum is from 358~d past
maximum and in that case we are able to measure the velocities
accurately. Thus, the epoch range in which we successfully measure
nebular velocities of [\ion{Fe}{II}] $\lambda$7155 and [\ion{Ni}{II}]
$\lambda$7378 is 115--358~d past maximum, with the majority of the
spectra in the range 160--358~d past maximum. Both \citet{Maeda10} and
\citet{Blondin12} had a similar range (\about150--400~d past maximum),
with the vast majority of their spectra being older than \about200~d past
maximum.

\subsection{Calculating Nebular Velocities}\label{ss:calc_v_neb}

Following \citet{Maeda10} and \citet{Blondin12}, we calculate nebular
velocities from the mean of the individual velocities of the
[\ion{Fe}{II}] $\lambda$7155 and [\ion{Ni}{II}] $\lambda$7378
features. Thus, the nebular velocity represents the typical velocity
of material synthesized at the earliest times in the SN explosion. 
Previous studies used the difference between the two
velocities as the uncertainty of the nebular velocity, but we use the
difference divided by two (since that is how much the average
differs from each measurement).

For the 4 SNe~Ia in which we are able to measure both
velocities in multiple late-time spectra, the nebular velocity is the
average of all [\ion{Fe}{II}] $\lambda$7155 and [\ion{Ni}{II}]
$\lambda$7378 velocities and the uncertainty is the average
difference. The calculated nebular velocities (\vneb) of
all 15 SNe~Ia with [\ion{Fe}{II}] $\lambda$7155 and [\ion{Ni}{II}]
$\lambda$7378 velocity measurements are listed in the ninth column of 
Table~\ref{t:final_vel}, along with various classifications and other
measured properties.

\setlength{\tabcolsep}{0.05in}
\begin{table*}
\scriptsize
\begin{center}
\caption{Summary of objects studied. Uncertainties are given in parentheses.}\label{t:final_vel}
\begin{tabular}{lcccccrrrcc}
\hline\hline
SN Name & $\Delta m_{15}(B)$$^\textrm{a}$ & \bvmax$^\textrm{a}$ & Wang & Si~II $\lambda$6355 Vel.$^\textrm{c}$ & Benetti$^{\phantom{\textrm{dj}}}$ & \multicolumn{1}{c}{$\dot{v}$$^\textrm{e}$} & \multicolumn{1}{c}{$v_0$$^\textrm{f}$} & \multicolumn{1}{c}{Nebular Vel.$^\textrm{g}$} & EW(Na) & EW(Ca) \\
     & (mag) & (mag) & Type$^\textrm{b}$ & ($10^3$~\kms) & Type$^{\textrm{d}\phantom{\textrm{j}}}$ & \multicolumn{1}{c}{(\kms~d$^{-1}$)} & \multicolumn{1}{c}{($10^3$~\kms)} & \multicolumn{1}{c}{(\kms)} & (\AA) & (\AA) \\
\hline

SN 1989M            &  1.10 (0.20)$\phantom{^\textrm{k}}$ &                           $\cdots$$\phantom{^\textrm{k}}$ &     HV$\phantom{^\textrm{i}}$ &             12.46 (0.10)$\phantom{^\textrm{h}}$ &            HVG$\phantom{^\textrm{hj}}$ &                         291.84 (139.49)$\phantom{^\textrm{hj}}$ &                13.19 (0.42)$\phantom{^\textrm{hj}}$ &     $560$ $\phantom{0}$(110) & 1.88 (0.73) & 0.36 (0.18) \\
SN 1993Z            &     $\cdots$$\phantom{^\textrm{k}}$ &                           $\cdots$$\phantom{^\textrm{k}}$ & $\cdots\phantom{^\textrm{i}}$ &                 $\cdots$$\phantom{^\textrm{h}}$ &       $\cdots$$\phantom{^\textrm{hj}}$ &              \multicolumn{1}{c}{$\cdots\phantom{^\textrm{hj}}$} &  \multicolumn{1}{c}{$\cdots\phantom{^\textrm{hj}}$} &    $1990$ $\phantom{0}$(630) & 1.56 (0.40) & 0.96 (1.67) \\
SN 1994D$^\dagger$  &  1.36 (0.05)$\phantom{^\textrm{k}}$ &                   $-0.053$ (0.035)$\phantom{^\textrm{k}}$ &      N$\phantom{^\textrm{i}}$ &             10.72 (0.10)$\phantom{^\textrm{h}}$ &            LVG$\phantom{^\textrm{hj}}$ &              33.43 $\phantom{00}$(6.77)$\phantom{^\textrm{hj}}$ &                10.60 (0.05)$\phantom{^\textrm{hj}}$ &   $-1720$ $\phantom{0}$(570) & 0.40 (0.44) & 0.09 (0.08) \\
SN 1998bu           &  1.05 (0.03)$\phantom{^\textrm{k}}$ &       $\phantom{-}$$0.292$ (0.032)$\phantom{^\textrm{k}}$ &      N$\phantom{^\textrm{i}}$ &                       10.84 (0.11)$^\textrm{h}$ & LVG$^{\textrm{h}\phantom{\textrm{j}}}$ &   46.52 $\phantom{00}$(5.38)$^{\textrm{h}\phantom{\textrm{j}}}$ &     11.01 (0.05)$^{\textrm{h}\phantom{\textrm{j}}}$ &   $-1800$ $\phantom{0}$(280) & 1.52 (1.12) & 0.39 (0.19) \\
SN 1999aa$^\dagger$ &  0.79 (0.05)$\phantom{^\textrm{k}}$ &                   $-0.093$ (0.033)$\phantom{^\textrm{k}}$ &           $\cdots^\textrm{i}$ &             10.50 (0.10)$\phantom{^\textrm{h}}$ & LVG$^{\textrm{j}\phantom{\textrm{h}}}$ &   14.33 $\phantom{00}$(3.52)$^{\textrm{j}\phantom{\textrm{h}}}$ &     10.49 (0.03)$^{\textrm{j}\phantom{\textrm{h}}}$ &     $760$ $\phantom{00}$(70) & 0.44 (0.64) & 0.27 (0.14) \\
SN 1999gh$^\dagger$ &  1.69 (0.05)$\phantom{^\textrm{k}}$ &       $\phantom{-}$$0.190$ (0.000)$\phantom{^\textrm{k}}$ &      N$\phantom{^\textrm{i}}$ &             11.02 (0.10)$\phantom{^\textrm{h}}$ &          FAINT$\phantom{^\textrm{hj}}$ &              46.49 $\phantom{0}$(10.13)$\phantom{^\textrm{hj}}$ &                11.20 (0.13)$\phantom{^\textrm{hj}}$ &    $1570$ $\phantom{0}$(280) & 0.50 (0.25) &    $\cdots$ \\
SN 2002bo           &  1.15 (0.04)$\phantom{^\textrm{k}}$ &       $\phantom{-}$$0.315$ (0.052)$\phantom{^\textrm{k}}$ &     HV$\phantom{^\textrm{i}}$ &             13.88 (0.10)$\phantom{^\textrm{h}}$ &            HVG$\phantom{^\textrm{hj}}$ &             245.14 $\phantom{00}$(8.12)$\phantom{^\textrm{hj}}$ &                13.61 (0.09)$\phantom{^\textrm{hj}}$ &                 $2430$ (100) & 2.36 (0.53) & 0.57 (0.31) \\
SN 2002fk$^\dagger$ &  1.20 (0.03)$\phantom{^\textrm{k}}$ &                   $-0.155$ (0.032)$\phantom{^\textrm{k}}$ &      N$\phantom{^\textrm{i}}$ & $\phantom{0}$9.51 (0.10)$\phantom{^\textrm{h}}$ & LVG$^{\textrm{j}\phantom{\textrm{h}}}$ &   11.21 $\phantom{00}$(8.28)$^{\textrm{j}\phantom{\textrm{h}}}$ &      9.74 (0.05)$^{\textrm{j}\phantom{\textrm{h}}}$ &   $-2060$ $\phantom{0}$(690) & 0.46 (0.09) & 0.26 (0.07) \\
SN 2003gs           &            1.83 (0.02)$^\textrm{k}$ &                 $\phantom{-}$$0.634$ (0.021)$^\textrm{k}$ & $\cdots\phantom{^\textrm{i}}$ &                       11.00 (0.20)$^\textrm{k}$ &       $\cdots$$\phantom{^\textrm{hj}}$ &              \multicolumn{1}{c}{$\cdots\phantom{^\textrm{hj}}$} &  \multicolumn{1}{c}{$\cdots\phantom{^\textrm{hj}}$} &                $2540$ (1330) & 0.98 (0.49) &    $\cdots$ \\
SN 2004dt           &  1.29 (0.05)$\phantom{^\textrm{k}}$ &                   $-0.080$ (0.034)$\phantom{^\textrm{k}}$ &     HV$\phantom{^\textrm{i}}$ &             14.34 (0.10)$\phantom{^\textrm{h}}$ &            HVG$\phantom{^\textrm{hj}}$ &             269.45 $\phantom{00}$(8.34)$\phantom{^\textrm{hj}}$ &                14.70 (0.10)$\phantom{^\textrm{hj}}$ &               $-2850$ (1380) & 1.80 (1.04) &    $\cdots$ \\
SN 2005cf$^\dagger$ &  1.08 (0.03)$\phantom{^\textrm{k}}$ &                   $-0.062$ (0.031)$\phantom{^\textrm{k}}$ &      N$\phantom{^\textrm{i}}$ &             10.26 (0.10)$\phantom{^\textrm{h}}$ &            HVG$\phantom{^\textrm{hj}}$ &                         106.69 (149.59)$\phantom{^\textrm{hj}}$ &                10.13 (0.25)$\phantom{^\textrm{hj}}$ &     $910$ $\phantom{0}$(500) & 0.14 (0.09) & 0.53 (0.26) \\
SN 2006X            &  1.10 (0.04)$\phantom{^\textrm{k}}$ &       $\phantom{-}$$1.260$ (0.034)$\phantom{^\textrm{k}}$ &     HV$\phantom{^\textrm{i}}$ &             15.26 (0.10)$\phantom{^\textrm{h}}$ & HVG$^{\textrm{j}\phantom{\textrm{h}}}$ &  184.80 $\phantom{00}$(5.82)$^{\textrm{j}\phantom{\textrm{h}}}$ &     15.63 (0.06)$^{\textrm{j}\phantom{\textrm{h}}}$ &    $2700$ $\phantom{0}$(410) & 2.04 (1.61) & 1.05 (0.46) \\
SN 2007le           &  1.02 (0.04)$\phantom{^\textrm{k}}$ &       $\phantom{-}$$0.280$ (0.039)$\phantom{^\textrm{k}}$ &     HV$\phantom{^\textrm{i}}$ &             12.09 (0.10)$\phantom{^\textrm{h}}$ &            HVG$\phantom{^\textrm{hj}}$ &              93.35 $\phantom{0}$(12.65)$\phantom{^\textrm{hj}}$ &                12.77 (0.18)$\phantom{^\textrm{hj}}$ &    $2030$ $\phantom{0}$(140) & 2.36 (0.62) & 0.33 (0.17) \\
SN 2008Q$^\dagger$  &  1.25 (0.08)$\phantom{^\textrm{k}}$ &                   $-0.012$ (0.031)$\phantom{^\textrm{k}}$ &     HV$\phantom{^\textrm{i}}$ &             11.09 (0.10)$\phantom{^\textrm{h}}$ & HVG$^{\textrm{j}\phantom{\textrm{h}}}$ &   82.66 $\phantom{0}$(16.22)$^{\textrm{j}\phantom{\textrm{h}}}$ &     11.64 (0.05)$^{\textrm{j}\phantom{\textrm{h}}}$ &   $-2200$ $\phantom{0}$(750) & 0.44 (0.22) &    $\cdots$ \\
SN 2011by           &  1.14 (0.03)$\phantom{^\textrm{k}}$ &                   $-0.061$ (0.032)$\phantom{^\textrm{k}}$ &      N$\phantom{^\textrm{i}}$ &             10.35 (0.14)$\phantom{^\textrm{h}}$ &            LVG$\phantom{^\textrm{hj}}$ &              52.83 $\phantom{00}$(8.76)$\phantom{^\textrm{hj}}$ &                10.40 (0.08)$\phantom{^\textrm{hj}}$ &   $-1450$ $\phantom{0}$(170) & 0.66 (0.16) & 0.47 (0.41) \\
\hline \hline
\multicolumn{10}{l}{$^\dagger$Object is part of the low-extinction sample (see text for details).} \\
\multicolumn{10}{l}{$^\textrm{a}$Values taken from \citet{Ganeshalingam10:phot_paper}.} \\
\multicolumn{10}{p{6.5in}}{$^\textrm{b}$Classification based on the velocity of the Si~II $\lambda$6355 line near maximum brightness \citep{Wang09}. `HV' = high velocity; `N' = normal. Taken from \citet{Silverman12:BSNIPII}.} \\
\multicolumn{10}{l}{$^\textrm{c}$Velocity of the Si~II $\lambda$6355 feature within 8~d of maximum brightness. Taken from \citet{Silverman12:BSNIPII} unless otherwise noted.} \\
\multicolumn{10}{p{6.5in}}{$^\textrm{d}$Classification based on the velocity gradient of the Si~II $\lambda$6355 line \citep{Benetti05}. `HVG' = high velocity gradient; `LVG' = low velocity gradient; `FAINT' = faint/underluminous. Taken from \citet{Silverman12:BSNIPII} unless otherwise noted.} \\
\multicolumn{10}{p{6.5in}}{$^\textrm{e}$Change in velocity with time, near maximum brightness (i.e., the velocity gradient). Taken from \citet{Silverman12:BSNIPII} unless otherwise noted.} \\
\multicolumn{10}{p{6.5in}}{$^\textrm{f}$Velocity of the Si~II $\lambda$6355 feature extrapolated to $t = 0$~d based on the velocity gradient. Taken from \citet{Silverman12:BSNIPII} unless otherwise noted.} \\
\multicolumn{10}{p{6.5in}}{$^\textrm{g}$Average nebular velocity of the [Fe~II] $\lambda$7155 and [Ni~II] $\lambda$7378 features for all late-time spectra of a given object. The uncertainty is the average difference between the velocity of the two features.} \\
\multicolumn{10}{p{6.5in}}{$^\textrm{h}$Si~II $\lambda$6355 velocity, Benetti type, velocity gradient, and $v_0$ measured from data presented by \citet{Matheson08}.} \\
\multicolumn{10}{l}{$^\textrm{i}$Classification scheme only applies to spectroscopically normal SNe~Ia.} \\
\multicolumn{10}{p{6.5in}}{$^\textrm{j}$Benetti type, velocity gradient, and $v_0$ measured from BSNIP data as well as data presented by \citet{Blondin12}.} \\
\multicolumn{10}{p{6.5in}}{$^\textrm{k}$$\Delta m_{15}(B)$ and \bvmax\ taken from \citet{Krisciunas09}. Si~II $\lambda$6355 velocity taken from \citet{SN2003gs_disc}.} \\
\hline\hline
\end{tabular}
\end{center}
\end{table*}
\normalsize
\setlength{\tabcolsep}{6pt}

The second column of Table~\ref{t:final_vel} gives the decline rate of
the light curve, $\Delta m_{15}(B)$, and the third column shows the
observed $B-V$ colour of the SN at $B$-band maximum brightness,
\bvmax, both taken from
\citet{Ganeshalingam10:phot_paper}. Also displayed in the table is the
velocity of the \ion{Si}{II} $\lambda$6355 feature within 8~d of
maximum brightness (fifth column) and a classification based on that
velocity of either normal velocity or high velocity \citep[fourth 
column; e.g.,][]{Wang09}. The rate of decrease of the expansion
velocity, also known as the velocity gradient ($\dot{v}$, seventh 
column), and the velocity of the \ion{Si}{II} $\lambda$6355 feature
extrapolated to $t = 0$~d using $\dot{v}$ ($v_0$, eighth column) are
also shown in Table~\ref{t:final_vel} along with a classification based on $\dot{v}$
\citep[sixth column; e.g.,][]{Benetti05}. The tenth (eleventh) column of
the table is the EW of narrow \ion{Na}{I}~D (\ion{Ca}{II}~H\&K)
absorption from the host galaxy of each SN (see
Section~\ref{sss:vneb_spec}
for more information on these measurements). Objects listed with a
`$\dagger$' are part of the low-extinction sample (see
Section~\ref{sss:vneb_phot}). Note that while most of the measurements come
directly from \citet{Silverman12:BSNIPII}, some of the near-maximum-light
\ion{Si}{II} $\lambda$6355 velocities were measured from data
presented by \citet{Matheson08} and \citet{Blondin12}.

The range of nebular velocities calculated in this study is
approximately $\pm 2900$~\kms, similar to the range found in
previous work \citep{Maeda10,Blondin12}. A comparison of the nebular
velocities of the 6 SNe~Ia that are found in the current sample as
well as previous studies shows that the values calculated herein are 
consistent with previous work, with average absolute differences of a
few hundred \kms\ and no detectable bias
\citep{Maeda10,Maeda11,Blondin12}. The average uncertainty of \vneb\
(which is half the difference between the individual velocities of the
[\ion{Fe}{II}] $\lambda$7155 and [\ion{Ni}{II}] $\lambda$7378
features) is \about490~\kms, slightly larger than the typical 
uncertainty in the measurements of \citet{Maeda10}.

\section{Analysis}\label{s:analysis}

\subsection{[\ion{Fe}{III}] $\lambda$4701}\label{ss:fe4701}

The FWHM of the [\ion{Fe}{III}] $\lambda$4701 feature has been seen to
possibly be anticorrelated with $\Delta m_{15}(B)$
\citep{Mazzali98}. In that work, both overluminous SN~1991T-like
objects {\it and} underluminous SN~1991bg-like objects were found to
follow the correlation. More recently, \citet{Blondin12} found a similar
result with a Pearson correlation coefficient of $-0.71$ for all of their
late-time data, but when their two underluminous SNe~Ia were removed
it decreased to $-0.17$ (i.e., no correlation). This latter result was
questioned, however, due to the fact that spectra at epochs earlier
than 250--300~d past maximum brightness were used \citep{Mazzali12}. 

\begin{figure}
\centering
\includegraphics[width=3.4in]{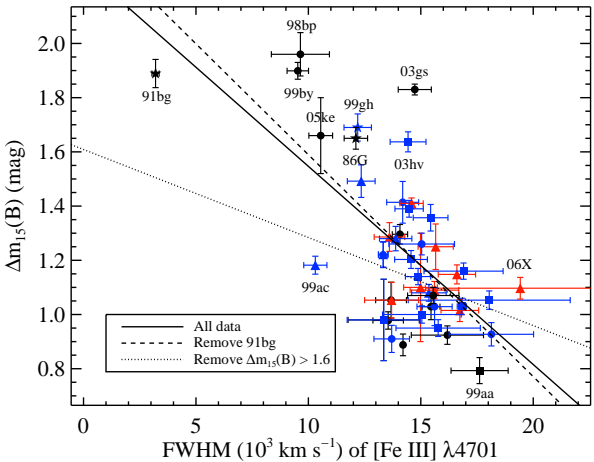} 
\caption{FWHM of [\ion{Fe}{III}] $\lambda$4701 versus $\Delta
  m_{15}(B)$ from BSNIP and 18 objects from \citet{Maeda11} and
  \citet{Blondin12}. Squares are LVG objects, triangles are HVG
  objects, 
  stars are FAINT objects, and circles are objects for which we are
  unable to determine a 
  Benetti Type. Blue points are normal-velocity objects, red points
  are HV objects, and black points are objects for which we are unable
  to determine a Wang
  Type. Some of the outlying points are labeled. Linear fits to the
  data are also displayed: all data (solid), 
  after removing SN~1991bg (dashed), and after removing all 6 objects
  with $\Delta m_{15}(B) > 1.6$~mag (dotted).}\label{f:fwhm4701}
\end{figure}

Figure~\ref{f:fwhm4701} shows the FWHM measurements of the
[\ion{Fe}{III}] $\lambda$4701 feature versus $\Delta m_{15}(B)$ from
the BSNIP spectra, as well as previously published values for 18 SNe
\citep{Maeda11,Blondin12}. The shape of each data point signifies its  
``Benetti Type'' (based on $\dot{v}$): squares are low velocity
gradient (LVG) objects, triangles are high velocity gradient (HVG)
objects, stars are FAINT objects, and circles are 
objects for which we are unable to determine a Benetti Type. The
colour of each point represents its ``Wang Type'' (based on
\ion{Si}{II} $\lambda$6355 velocity): blue is for normal-velocity
objects, red is 
for HV objects, and black is objects for which we are unable to
determine a Wang Type. Some of the outlying points are labeled with
their object name.

As in previous studies \citep{Mazzali98,Blondin12}, the BSNIP data show
that the FWHM of [\ion{Fe}{III}] $\lambda$4701 is 
anticorrelated with $\Delta m_{15}(B)$ (Pearson correlation
coefficient of $-0.66$) and the linear fit to all of the data is shown
as the solid line in Figure~\ref{f:fwhm4701}. The slope of this line,
$-0.073$~mag/($10^3$~\kms), is consistent with that of
\citet{Blondin12} but inconsistent with that of \citet{Mazzali98}, and
the correlation is significant at the $>3\sigma$ level. If we remove
only the underluminous SN~1991bg, the linear fit is nearly unchanged
(the dashed line in the figure) and the correlation weakens only
slightly (Pearson coefficient of $-0.61$). If we instead remove the 8
objects with $\Delta m_{15}(B) > 1.6$~mag (i.e., all of the
underluminous objects in the dataset), the correlation effectively
disappears (Pearson coefficient of $-0.34$) and the slope of the
linear fit (the dotted line in Figure~\ref{f:fwhm4701}) is consistent
with 0 at the \about2$\sigma$ level.

The findings of \citet{Blondin12} confirmed above have recently been
questioned due to their use of spectra at epochs earlier than
250--300~d past maximum brightness \citep{Mazzali12}. To investigate
this possibility, we performed a separate analysis of the FWHM of the
[\ion{Fe}{III}] $\lambda$4701 feature versus $\Delta m_{15}(B)$ using
only the oldest spectra from BSNIP, \citet{Maeda11}, and
\citet{Blondin12}. There were 16 (8) objects with spectra obtained at
$t >$~250~d (300~d) past maximum brightness, and only a modest
anticorrelation was found for these data with a Pearson
coefficient of $-0.55$ ($-0.69$). When removing the 2 (1)
underluminous objects with $\Delta m_{15}(B) > 1.6$~mag having 
spectra at these latest epochs, the possible 
correlation completely disappears. Thus, it seems that at all epochs
the possible relationship between the FWHM of the [\ion{Fe}{III}]
$\lambda$4701 feature and $\Delta m_{15}(B)$ is being driven almost
completely by the most underluminous SNe~Ia, perhaps indicating the
existence of two groups of points rather than a linear correlation.

\begin{figure}
\centering
\includegraphics[width=3.4in]{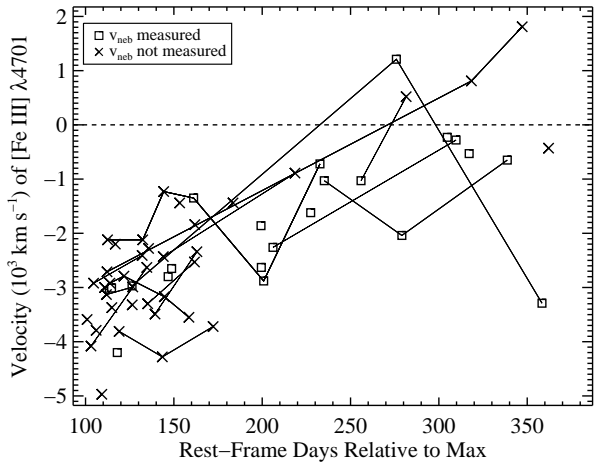} 
\caption{All velocities of [\ion{Fe}{III}] $\lambda$4701 measured in
  this work. Solid lines connect multiple velocities of individual
  objects. Squares represent spectra for which we measured a nebular
  velocity; $\times$ symbols represent spectra for which we did
  not.}\label{f:v4701_t}
\end{figure}

 Regarding the velocity of the [\ion{Fe}{III}] $\lambda$4701 feature,
\citet{Stritzinger06} found no significant temporal change in their
measurements and \citet{Maeda10} state that the feature shows
``virtually no Doppler shift.'' However, later work from the latter
group show blueshifts of 2000--4500~\kms\ for $t < 200$~d, with the
velocity steadily approaching 0 for $t > 200$~d
\citep{Maeda10c,Maeda11}. This 
more recent result is also seen in the BSNIP data. In
Figure~\ref{f:v4701_t} all measured velocities of [\ion{Fe}{III}]
$\lambda$4701 are plotted, with solid lines connecting multiple
velocities of individual objects. Squares represent spectra for which
we measured a nebular velocity, while $\times$ symbols represent
spectra for which we did not.

The overall trend \citep[which matches what was found by][]{Maeda10c}
is clear: velocities are blueshifted typically by
\about2000--4000~\kms\ at $t = 
100$~d and the amount of blueshift decreases approximately linearly
with time. This temporal evolution is quite slow, however, with a
typical change in velocity of only \about10--20~\kms~d$^{-1}$. The
velocity of the [\ion{Fe}{III}] $\lambda$4701 feature is uncorrelated
with the nebular velocity and hence with the individual velocity of either
[\ion{Fe}{II}] $\lambda$7155 or [\ion{Ni}{II}]
$\lambda$7378. Furthermore, the [\ion{Fe}{III}] $\lambda$4701 velocity
is also uncorrelated with the \ion{Si}{II} $\lambda$6355 velocity and
$\dot{v}$.

\subsection{[\ion{Fe}{II}] $\lambda$7155 and [\ion{Ni}{II}] $\lambda$7378}\label{ss:fe7155}

The individual velocities of the [\ion{Fe}{II}] $\lambda$7155 and
[\ion{Ni}{II}] $\lambda$7378 features show a relatively large scatter
at all epochs, which is reflected in the large range of calculated
nebular velocities. The temporal velocity evolution of both features
is even slower than that of the [\ion{Fe}{III}] $\lambda$4701 feature,
with typical changes of a few \kms~d$^{-1}$.

Since the average of the velocities of [\ion{Fe}{II}] $\lambda$7155
and [\ion{Ni}{II}] $\lambda$7378 is used to calculate \vneb, it is
instructive to compare these two velocities with each other. The
difference in these velocities has a relatively large scatter at all
epochs, but we find no bias above or below zero velocity. The average
{\it absolute} difference between the two velocities, however, is
\about900~\kms, which is reflected in our \vneb\ uncertainties
discussed above. Despite this, the two velocities are highly
correlated (Pearson coefficient of 0.84) and the difference between the two 
has a typical temporal evolution of $< 10$~\kms~d$^{-1}$.

\subsection{Nebular Velocities}\label{ss:v_neb}

\subsubsection{Spectral Comparisons}\label{sss:vneb_spec}

Like the [\ion{Fe}{II}] $\lambda$7155 and [\ion{Ni}{II}] $\lambda$7378
velocities from which it is calculated, the nebular velocity does not
change much with time \citep[typical values being only a few
\kms~d$^{-1}$, similar to what has been seen
previously; e.g.,][]{Maeda11}. Furthermore, all four SNe~Ia which have
measured [\ion{Fe}{II}] $\lambda$7155 and [\ion{Ni}{II}] $\lambda$7378 
velocities in multiple late-time spectra show no significant change in
their nebular velocities with time (for $130 < t < 360$~d). Thus, a
single measurement of \vneb\ appears to be sufficient to determine the
typical nebular velocity of a given SN~Ia.

Measurements of the velocity gradient were made for all objects with
multiple near-maximum-brightness spectra in BSNIP, in \citet{Matheson08},
or in \citet{Blondin12} (i.e., all SNe~Ia listed in
Table~\ref{t:final_vel} except SNe~1993Z and 2003gs). The $\dot{v}$
values calculated in this work are consistent at the 2$\sigma$ level
with those found in previous studies \citep{Benetti05,Maeda10}. We
also plot previously published $\dot{v}$ and \vneb\ values for 12
SNe~Ia from \citet{Maeda11} and \citet{Blondin12}.

\begin{figure}
\centering
\includegraphics[width=3.4in]{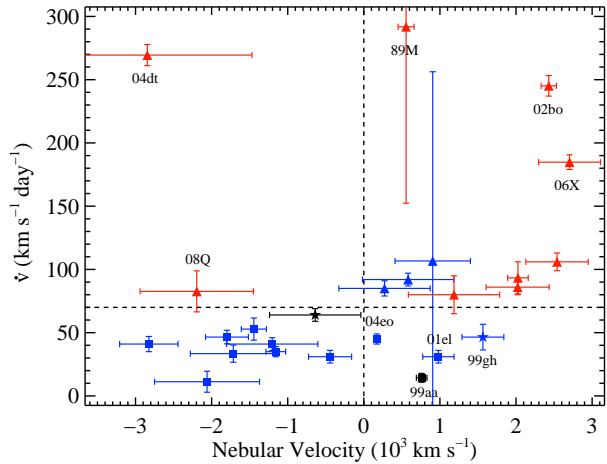} 
\caption{Nebular velocity versus velocity gradient. Shapes and colours
  of the data points are the same as in
  Figure~\ref{f:fwhm4701}. The vertical dashed line at 0~\kms\
  separates blueshifted (left) and redshifted (right) nebular
  velocities. The horizontal dashed line at 70~\kms~d$^{-1}$ separates
  LVG (below) and HVG (above) objects.}\label{f:vneb_vdot} 
\end{figure}

Of the 25 objects with measured velocity gradients, 11 are LVG, 12 are
HVG, and 2 are FAINT, which is similar to the distribution of types
found by \citet{Maeda10}, though they have a slightly larger fraction
of LVG objects. Figure~\ref{f:vneb_vdot} presents these
velocity-gradient measurements versus the nebular velocity; shapes and
colours of the data points represent the Benetti and Wang types,
respectively, as in Figure~\ref{f:fwhm4701}. The vertical dashed line
at 0~\kms\ is the demarcation between blueshifted (left) and redshifted
(right) nebular velocities; the horizontal dashed line at
70~\kms~d$^{-1}$ is the demarcation between LVG (below) and HVG
(above) objects.

\citet{Maeda10} found a correlation between \vneb\ and $\dot{v}$, and
the BSNIP data support this conclusion, with some caveats. The Pearson
coefficient for all of the data in Figure~\ref{f:vneb_vdot} is only
0.25 and the Spearman rank correlation coefficient is
0.39.\footnote{This is perhaps the more appropriate value to examine here
  since the relationship appears to be nonlinear.} However, SN~2004dt 
was found to be somewhat peculiar \citep[showing evidence for
high-velocity and clumpy ejecta at early times; e.g.,][]{Altavilla07} and did
not follow the relationship of \citet{Maeda10}. Also, there are only two
BSNIP spectra of SN~1989M and they were obtained just 1~d apart, thus
the lever arm for the $\dot{v}$ calculation is small and the accuracy
of the measurement is small (hence the large error bars for that data
point).

Removing SN~2004dt (SNe~2004dt and 1989M) yields a Pearson
correlation coefficient of 0.51 (0.62) and a Spearman rank correlation
coefficient of 0.56 (0.59). This indicates that the nebular velocity
is correlated with the velocity gradient, with the significance of the
correlation at the 2$\sigma$ (3$\sigma$) level.
If we ignore SN~2004dt, all HVG objects have redshifted
nebular lines, except SN~2008Q which is extremely close to the LVG/HVG
border. We also find that nearly 3/4 of all LVG objects have
blueshifted nebular spectral features. A Kolmogorov-Smirnov (KS) test
on the nebular velocities of LVG and HVG objects implies that they
very likely come from different parent populations ($p \approx
0.016$). When removing SN~2004dt, the difference is even stronger ($p 
\approx 0.007$).

\begin{figure}
\centering
\includegraphics[width=3.4in]{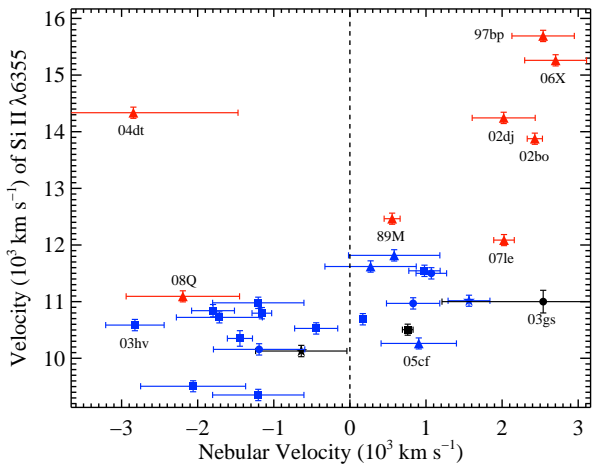} 
\caption{Nebular velocity versus \ion{Si}{II} $\lambda$6355
  velocity. Shapes and colours 
  of the data points are the same as in
  Figure~\ref{f:fwhm4701}. Some of the outlying points are
  labeled. The vertical dashed line at 0~\kms\ 
  separates blueshifted (left) and redshifted (right) nebular
  velocities.}\label{f:vneb_v} 
\end{figure}

The near-maximum-brightness \ion{Si}{II} $\lambda$6355 velocity can
also be compared to the nebular velocity for 14 of the SNe~Ia in
Table~\ref{t:final_vel}, as well as 15 SNe~Ia from \citet{Maeda11} and
\citet{Blondin12}. We measure the \ion{Si}{II} $\lambda$6355 velocity
only in BSNIP spectra within 8~d of maximum
brightness.\footnote{Spectra within 5~d of maximum  
brightness were used by \citet{Silverman12:BSNIPII}, but applying
this constraint would remove three objects from the study of \ion{Si}{II}
$\lambda$6355 velocity versus
nebular velocity.} When multiple BSNIP spectra within 8~d of
maximum exist for a given object, only the one closest to maximum
brightness is used for the \ion{Si}{II} $\lambda$6355 velocity
measurement. The near-maximum-light
velocities are plotted against the nebular velocities in 
Figure~\ref{f:vneb_v} with shapes and colours of data points the
same as in Figure~\ref{f:fwhm4701}; the vertical dashed line separates
blueshifted (left) and redshifted (right) nebular velocities. 

A moderate correlation exists for all of the data presented in
Figure~\ref{f:vneb_v} (Pearson and Spearman coefficients are both
0.51). If SN~2004dt is again ignored, a strong correlation exists
(Pearson coefficient of 0.70 and Spearman coefficient of 0.64). Most
HV objects have redshifted emission lines at late time, but
normal-velocity objects have both blueshifted {\it and} redshifted
nebular features. A KS test of the nebular velocities indicates,
however, that the HV and normal-velocity objects do come from
different parent populations ($p \approx 0.021$ for all of the data and
$p \approx 0.009$ when ignoring SN~2004dt).

The overall trend and even the location of each
object in the parameter space (given the measurement uncertainties) is
quite similar to Figure~\ref{f:vneb_vdot}.  This is generally
unsurprising given the results above involving the velocity gradient,
since most (but not all) HV (normal-velocity) objects are also HVG
(LVG) objects \citep[Table~\ref{t:final_vel}
of][]{Wang06,Silverman12:BSNIPII}. Furthermore, the \ion{Si}{II}
$\lambda$6355 velocity at $t = 0$~d (i.e., $v_0$) can be approximated
from the velocity gradient and, as expected, these velocities at
maximum brightness correlate in almost exactly the same manner as what
is seen in Figure~\ref{f:vneb_v}.

To investigate whether late-time observations of SNe~Ia are related to
the amount of circumstellar interaction a SN~Ia undergoes,
\citet{Forster12} measured the EW of narrow absorption from
\ion{Na}{I}~D and \ion{Ca}{II}~H\&K in the host galaxies of SNe~Ia and
compared them to nebular velocities. Using 212 BSNIP spectra of the 15
objects listed in Table~\ref{t:final_vel}, we attempt to measure the
EW of host-galaxy \ion{Na}{I}~D and \ion{Ca}{II}~H\&K absorption. Even
in the relatively low-resolution spectra of the BSNIP sample, we are
able to measure \ion{Na}{I}~D absorption in all 15 SNe~Ia and
\ion{Ca}{II}~H\&K absorption in 11 of them. The final EWs listed in
the last two columns of Table~\ref{t:final_vel} are the median of all
EW measurements for each object (and the uncertainty is the standard
deviation).

There are 6 objects in the current work that were also studied by 
\citet{Forster12}, and the EW measurements for all of these SNe~Ia are
consistent at the $2\sigma$ level. In contrast to previous work
\citep{Forster12}, the BSNIP data do {\it not} show a significant
correlation between the EWs of \ion{Na}{I}~D and \ion{Ca}{II}~H\&K
absorption. \citet{Forster12} also found, at quite high significance,
that objects with redshifted nebular velocities have larger EWs of
\ion{Na}{I}~D and \ion{Ca}{II}~H\&K absorption. They interpret this
result as indicating that these objects have stronger circumstellar
interaction or larger amounts of circumstellar material, and thus 
they claim that nebular velocities are directly related to SN~Ia
environments.

The objects with redshifted \vneb\ in BSNIP {\it tend}
to have stronger absorption from \ion{Na}{I}~D and \ion{Ca}{II}~H\&K,
but this is not always the case, and the difference between objects with
blueshifted and redshifted nebular velocities is not
significant. Therefore, we are unable to confirm the findings of
\citet{Forster12} that the environments of SNe~Ia directly affect
nebular velocities (or vice versa).

\subsubsection{Photometric Comparisons}\label{sss:vneb_phot}

\begin{figure}
\centering
\includegraphics[width=3.4in]{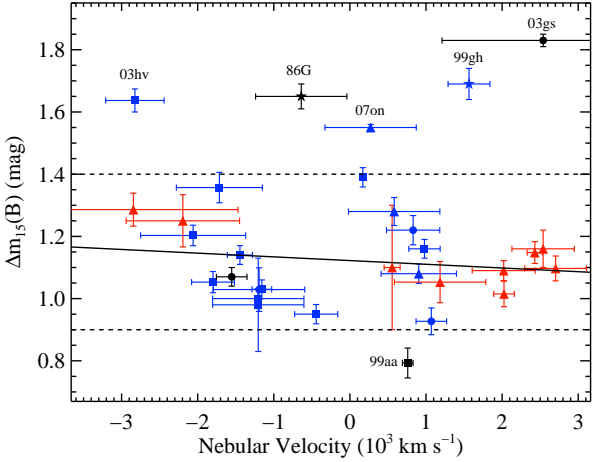} 
\caption{Nebular velocity versus $\Delta m_{15}(B)$. Shapes and
  colours of the data points are the same as in
  Figure~\ref{f:fwhm4701}. The six objects 
  with $\Delta m_{15}(B)$ values which fall outside the ``normal''
  range ($0.9 < \Delta m_{15}(B) < 1.4$~mag, denoted by the
  horizontal dashed lines) are labeled. The solid line is the linear
  fit to the data when ignoring these seven
  outliers.}\label{f:vneb_dm15}
\end{figure}

As mentioned above, the photometric parameters used in this work can
be found in \citet{Ganeshalingam10:phot_paper}. Of the 15 objects
listed in Table~\ref{t:final_vel}, 14 have near-maximum-brightness
light curves which allow for the measurement of $\Delta 
m_{15}(B)$. We also analyse 17 additional objects from \citet{Maeda11}
and \citet{Blondin12}. Most of them have $0.9 < \Delta m_{15}(B) <
1.4$~mag, which represents the range of ``normal'' SNe~Ia
\citep[e.g.,][]{Ganeshalingam10:phot_paper}. This range is denoted by
the horizontal dashed lines in Figure~\ref{f:vneb_dm15}, where we plot
\vneb\ against $\Delta m_{15}(B)$ and have labeled the six objects
that fall outside the normal range of $\Delta m_{15}(B)$ values.

\begin{figure}
\centering
\includegraphics[width=3.4in]{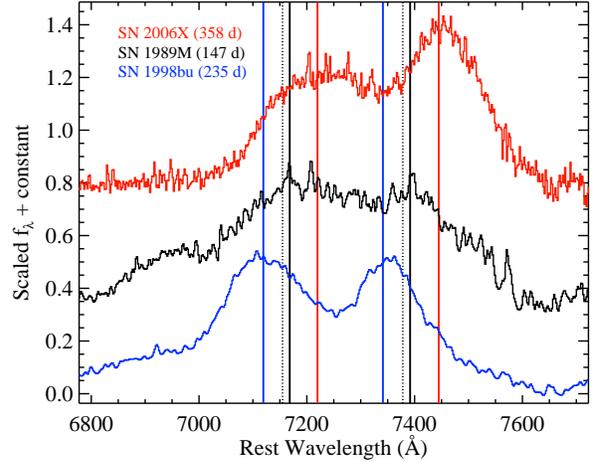} 
\caption{The deredshifted [\ion{Fe}{II}] $\lambda$7155 and
  [\ion{Ni}{II}] $\lambda$7378 features of SNe~2006X, 1989M, and
  1998bu, all of which have $\Delta m_{15}(B) \approx 1.1$~mag. The solid
  vertical lines are the nebular velocity for both features of each
  object, while the dotted vertical lines are the rest wavelengths of
  the two features.}\label{f:comp} 
\end{figure}

When including all objects presented in the figure, the data are
completely uncorrelated (Pearson correlation coefficient of $-0.009$),
similar to what has been seen previously \citep{Blondin12}. Even when
the six outliers (SNe~1986G, 1999aa, 1999gh, 2003gs, 2003hv, and
2007on) are ignored, there is still no significant evidence of a
correlation between \vneb\ and $\Delta m_{15}(B)$ (Pearson correlation
coefficient of $-0.16$). The solid line in Figure~\ref{f:vneb_dm15} is
the linear fit to the data when ignoring these six
outliers. Furthermore, the slope of this linear fit is consistent with
zero, and it is clear that for a given $\Delta m_{15}(B)$ there is a
huge range of possible \vneb\ values. This result, which matches that
of previous studies \citep{Maeda10,Blondin12}, is illustrated in
Figure~\ref{f:comp}, where the [\ion{Fe}{II}] $\lambda$7155 and
[\ion{Ni}{II}] $\lambda$7378 features are plotted for three
normal-luminosity objects all with $\Delta m_{15}(B) \approx
1.1$~mag. The solid vertical lines represent the nebular velocity for
both features of each object while the dotted vertical lines denote
the rest wavelengths of the two features. Despite their nearly
identical light curves, these three objects show significantly
different nebular velocities with a range 
spanning \about4500~\kms. 

\begin{figure}
\centering
\includegraphics[width=3.4in]{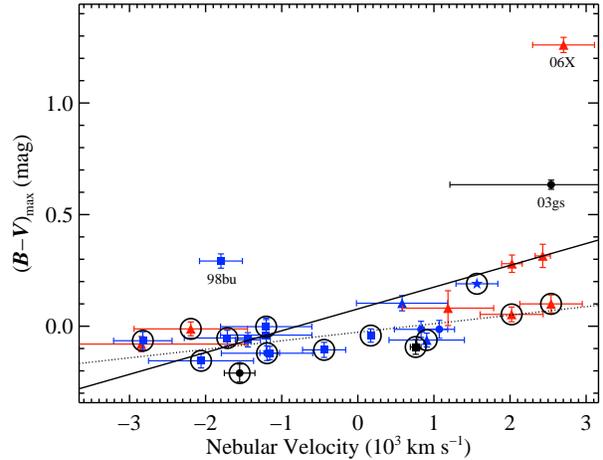} 
\caption{Nebular velocity versus observed \bvmax. Shapes and colours
  of the data points are the same as in Figure~\ref{f:fwhm4701}. Two
  of the labeled outliers (SNe~1998bu and 2006X) have significant
  reddening from their host galaxies, while the third labeled outlier
  (SN~2003gs) shows little host-galaxy reddening. The solid line is
  the linear fit to all of the data. The circled points are objects in the
  low-extinction sample and the dotted line is the linear fit to this
  subset.}\label{f:vneb_bv}
\end{figure}

In \citet{Silverman12:BSNIPIII}, \bvmax\ was defined as the observed
$B-V$ colour of the SN at $B$-band maximum brightness. In
Figure~\ref{f:vneb_bv} we plot \vneb\ versus 
\bvmax\ using this definition for the BSNIP data\footnote{SNe~1989M and 
1993Z do not have colour information available.} as well as 14 additional
SNe~Ia from \citet{Maeda11} and \citet{Blondin12}. In this parameter
space there appears to be three outliers (labeled in
Figure~\ref{f:vneb_bv}). Two of these, SNe~1998bu and 2006X, are known
to have substantial extinction from their host galaxies
\citep[e.g.,][]{Wang08,Matheson08}. On the other hand, the third
outlier (SN~2003gs) shows no evidence for significant host-galaxy
extinction \citep{Krisciunas09}.

Utilising all data in Figure~\ref{f:vneb_bv}, \vneb\ is found to be
correlated (at the $>3\sigma$ level) with \bvmax\ (Pearson coefficient
of 0.59), and the linear fit to these data is shown in the figure as
the solid line. \citet{Blondin12} found no correlation between \vneb\
and {\it intrinsic} $B-V$ colour at $B$-band maximum for their full
sample or when removing the peculiar SN~2004dt. However, they see a
correlation similar to ours (with Pearson correlation coefficient of
0.82) for objects with $v_\textrm{neb} > -2000$~\kms; note that most
of the objects presented herein also have $v_\textrm{neb} >
-2000$~\kms.

Since \bvmax\ is an {\it observed} SN colour (and thus a mixture of
intrinsic SN colour and reddening from the host galaxy), we also
investigate possible correlations between \vneb\ and \bvmax\ for
various low-extinction subsets of our sample. Simply removing
SN~2006X or both SNe~2006X and 1998bu from the analysis (since they
are known to have large host-galaxy extinction) increases the Pearson
coefficient to 0.60 and 0.77, respectively (and both correlations are
significant at the \about3$\sigma$ level). Concentrating only on
objects with $\left(B-V\right)_{\rm max} < 0.2$~mag, we find a
similarly strong correlation (Pearson coefficient of 0.64) again at
the $>3\sigma$ level. 

On the other hand, a low-extinction sample was defined by
\citet{Blondin12}, and no correlation between \vneb\ and intrinsic
\bvmax\ was detected. In this work we use similar criteria as
\citet{Blondin12} to define our low-extinction sample. A SN~Ia is
considered part of this sample if it resides in an E/S0 host galaxy,
is found far (in projected distance) from a host galaxy of any type,
or has a low value of $A_V$ according to {\tt MLCS2k2}
\citep{Jha07}. Objects in the low-extinction sample are marked with a
`$\dagger$'  in Table~\ref{t:final_vel} and are circled in
Figure~\ref{f:vneb_bv}. In contrast to \citet{Blondin12}, we find a
correlation between \vneb\ and observed \bvmax\ for the low-extinction
sample (Pearson correlation coefficient of 0.63) at the
\about2.5$\sigma$ level, and the linear fit to this subset of SNe~Ia
is plotted in Figure~\ref{f:vneb_bv} as a dotted line.

\section{Conclusions}\label{s:conclusions}

In this work we have analysed late-time ($t > 100$~d) optical spectra
of low-redshift ($z < 0.1$) SNe~Ia which mostly come from the BSNIP
dataset \citep{Silverman12:BSNIPI}. The lone exception is SN~2011by,
and we present and analyse 9 spectra of this object (2 of which are at
late times). After removing objects that do not follow the Phillips
relation \citep{Phillips93} and spectra that do not cover either of
the wavelength ranges of interest (namely the [\ion{Fe}{III}]
$\lambda$4701, [\ion{Fe}{II}] $\lambda$7155, and [\ion{Ni}{II}]
$\lambda$7378 features), we are left with 34 SNe~Ia with 60 spectra
having $t > 100$~d. Of these, 7 were studied by \citet{Maeda10}, and
those same 7 plus 2 more were studied by \citet{Blondin12}. We add to
the BSNIP data nebular spectral feature measurements of 20 SNe~Ia from
previously published work \citep{Maeda11,Blondin12}. This
sample represents the largest set of late-time SN~Ia spectra ever
analysed.

We attempted to measure the FWHM and velocity of the three nebular
emission features mentioned above and are able to measure the
[\ion{Fe}{III}] $\lambda$4701 line in all late-time data discussed in this
work. On the other hand, we are only able to calculate velocities of
the [\ion{Fe}{II}] $\lambda$7155 and [\ion{Ni}{II}] $\lambda$7378
features in 22 spectra of 15 SNe~Ia. To successfully measure these
features it seems that the SN~Ia should be spectroscopically normal,
the spectrum should have S/N~$\ga 20$~px$^{-1}$, and the spectrum
should be older than \about160~d past maximum brightness. Following
\citet{Maeda10}, we calculate \vneb\ from the mean of the individual
velocities of the [\ion{Fe}{II}] $\lambda$7155 and [\ion{Ni}{II}]
$\lambda$7378 features.

The FWHM of [\ion{Fe}{III}] $\lambda$4701 is anticorrelated with
$\Delta m_{15}(B)$. However, if the 6 objects with $\Delta m_{15}(B) >
1.6$~mag are removed, the correlation effectively
disappears. The same results are found when using only spectra from
250--300~d past maximum brightness. Thus, this possible relationship
seems to be driven solely by the most underluminous SNe~Ia. This
feature is blueshifted by \about2000--4000~\kms\ at $t = 100$~d, and
the amount of blueshift decreases approximately linearly with
time. However, this temporal evolution is quite slow, with a typical
change in velocity of only \about10--20~\kms~d$^{-1}$.

The nebular velocity likewise does not change much with time (typical
values are a few~\kms~d$^{-1}$). Furthermore, the BSNIP data weakly
support the finding of \citet{Maeda10} that \vneb\ and $\dot{v}$ are
correlated. Nearly all HVG objects have redshifted nebular lines and
most LVG objects have blueshifted nebular spectral features. A similar
result is found when using the near-maximum-brightness velocity
instead of the velocity gradient and assuming HV (normal-velocity)
objects are also HVG (LVG) objects \citep[which is a reasonable
assumption, e.g.,][]{Silverman12:BSNIPII}. The data studied in this
work also show no significant correlation between the EWs of
\ion{Na}{I}~D or \ion{Ca}{II}~H\&K absorption and \vneb; thus, we are
unable to confirm previous claims that the environments of SNe~Ia
directly affect nebular velocities, or vice versa
\citep{Forster12}. No significant correlation is found between \vneb\ and
$\Delta m_{15}(B)$, and SNe~Ia with nearly identical light curves can
have a wide range of nebular velocities, spanning \about4500~\kms\ in
Figure~\ref{f:comp}. Our data also indicate a correlation between
observed \bvmax\ and \vneb.

Although the combined dataset analysed in this work constitutes the
largest late-time SN~Ia spectral dataset ever studied, it still
contains only a handful of objects. The analysis herein could
certainly be improved and extended by obtaining more nebular SN~Ia
spectra. While BSNIP \citep[and the CfA,][]{Blondin12} has gathered,
published, and analysed their SN~Ia spectral data, larger transient
searches such as the Palomar Transient Factory
\citep[PTF;][]{Rau09,Law09} and Pan-STARRS \citep{Kaiser02} will be
able to add significantly to the number of late-time SN~Ia spectra.

\section*{Acknowledgments}

We would like to thank G. Canalizo, S. B. Cenko, K. Clubb, M. Cooper,
A. Diamond-Stanic, E. Gates, K. Hiner, M. Kandrashoff, M. Lazarova,
A. Miller, P. Nugent, and the overall LAMP collaboration
\citep{Barth11} for their
help with the observations of SN~2011by. R.~J.~Foley and
S.~W.~Jha discussed earlier drafts of this work with us, 
and the anonymous referee provided comments and suggestions that
improved the manuscript. 
We are grateful to the staffs at the Lick and Keck Observatories for
their support. Some of the data utilised herein were obtained at the
W. M. Keck Observatory, which is operated as a scientific partnership
among the California Institute of Technology, the University of
California, and NASA; the observatory was made possible by the
generous financial 
support of the W. M. Keck Foundation. The authors wish to recognise
and acknowledge the very significant cultural role and reverence that
the summit of Mauna Kea has always had within the indigenous Hawaiian
community; we are most fortunate to have the opportunity to conduct
observations from this mountain.
This work is supported by NSF grants AST-0908886 and AST-1211916, DOE
grants DE-FC02-06ER41453 (SciDAC) and DE-FG02-08ER41563, the TABASGO
Foundation, and the Christopher R. Redlich Fund. J.M.S. is grateful to
Marc J. Staley for a Graduate Fellowship. KAIT and its ongoing
operation were made possible by donations from Sun Microsystems, Inc.,
the Hewlett-Packard Company, AutoScope Corporation, Lick Observatory,
the NSF, the University of California, the Sylvia \& Jim Katzman
Foundation, and the TABASGO Foundation. We dedicate this paper to
the memory of Wallace~L.~W.~Sargent (deceased 2012~Oct.~29); the
discovery of SN~1985F by \citet{Filippenko85} sparked A.V.F.'s
intense interest in supernovae and dramatically affected his career.



\appendix

\section{Observations of SN~2011by}\label{s:appendix}

SN~2011by was discovered by \citet{11by_disc} on 2011~Apr.~26.8 UT
in NGC~3972 with J2000.0 coordinates $\alpha = 11^{\mathrm{h}} 
55^{\mathrm{m}} 45.56^{\mathrm{s}}$, $\delta = +55^{\circ} 19\arcmin
33\farcs8$, and it was classified \about1~d after discovery as a young
SN~Ia by \citet{11by_class}. \citet{11by_xr1} detected X-ray emission
coincident with SN~2011by \about40~d after discovery using {\it
  Swift}/XRT, but the emission was found to be unassociated with the
SN using {\it Chandra} observations taken \about3~weeks later
\citep{11by_xr2}. 

\begin{figure}
\centering
\includegraphics[width=3.4in]{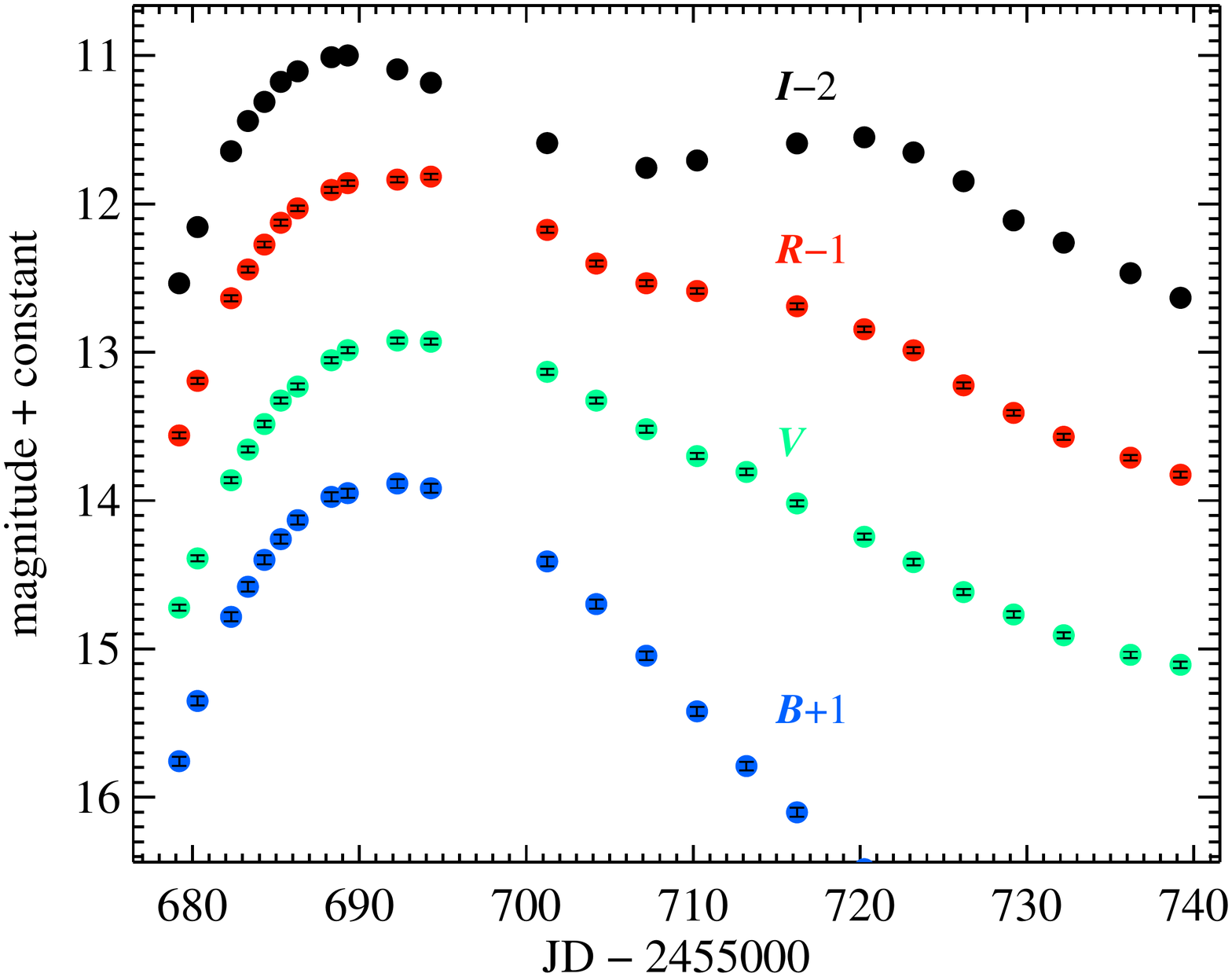}
\caption{\protect\hbox{$BV\!RI$} light curves of
  SN~2011by from KAIT.}\label{f:11by_lc}  
\end{figure}

Broad-band \bvri photometry of SN~2011by was obtained using the 0.76-m
Katzman Automatic Imaging Telescope (KAIT) at Lick Observatory
(\citealt{li00:kait,Filippenko01}) and is shown 
in Figure~\ref{f:11by_lc}. The
data were reduced using standard 
procedures, the details of which can be found in
\cite{Ganeshalingam10:phot_paper}. The optical light curves of
SN~2011by indicate that it reached $B$-band maximum brightness on
2011~May~9.9 UT and that it had $\Delta m_{15}(B) = 1.14 \pm 0.03$~mag.

\begin{figure}
\centering$
\begin{array}{c}
\includegraphics[width=3.4in]{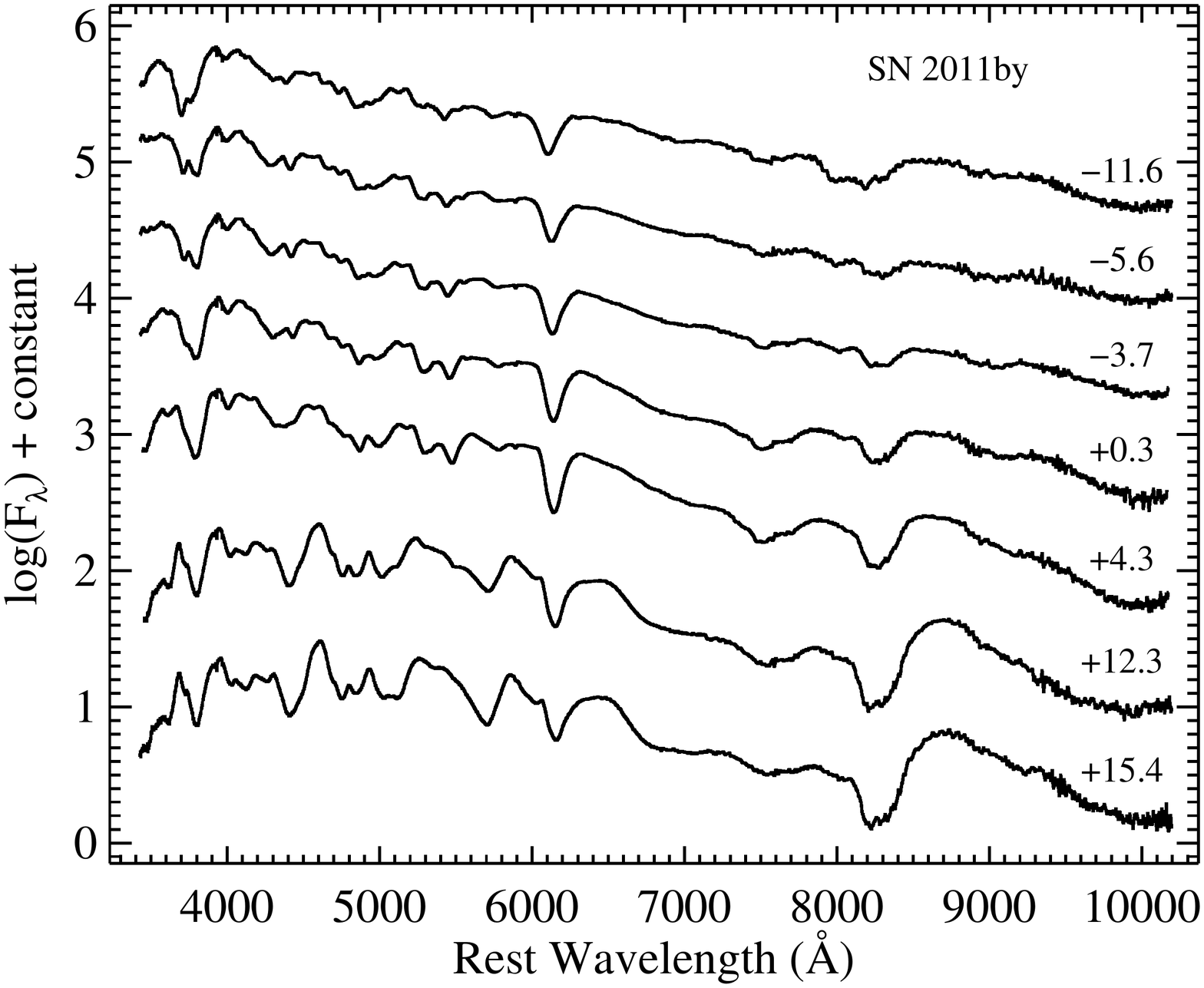} \\
\includegraphics[width=3.4in]{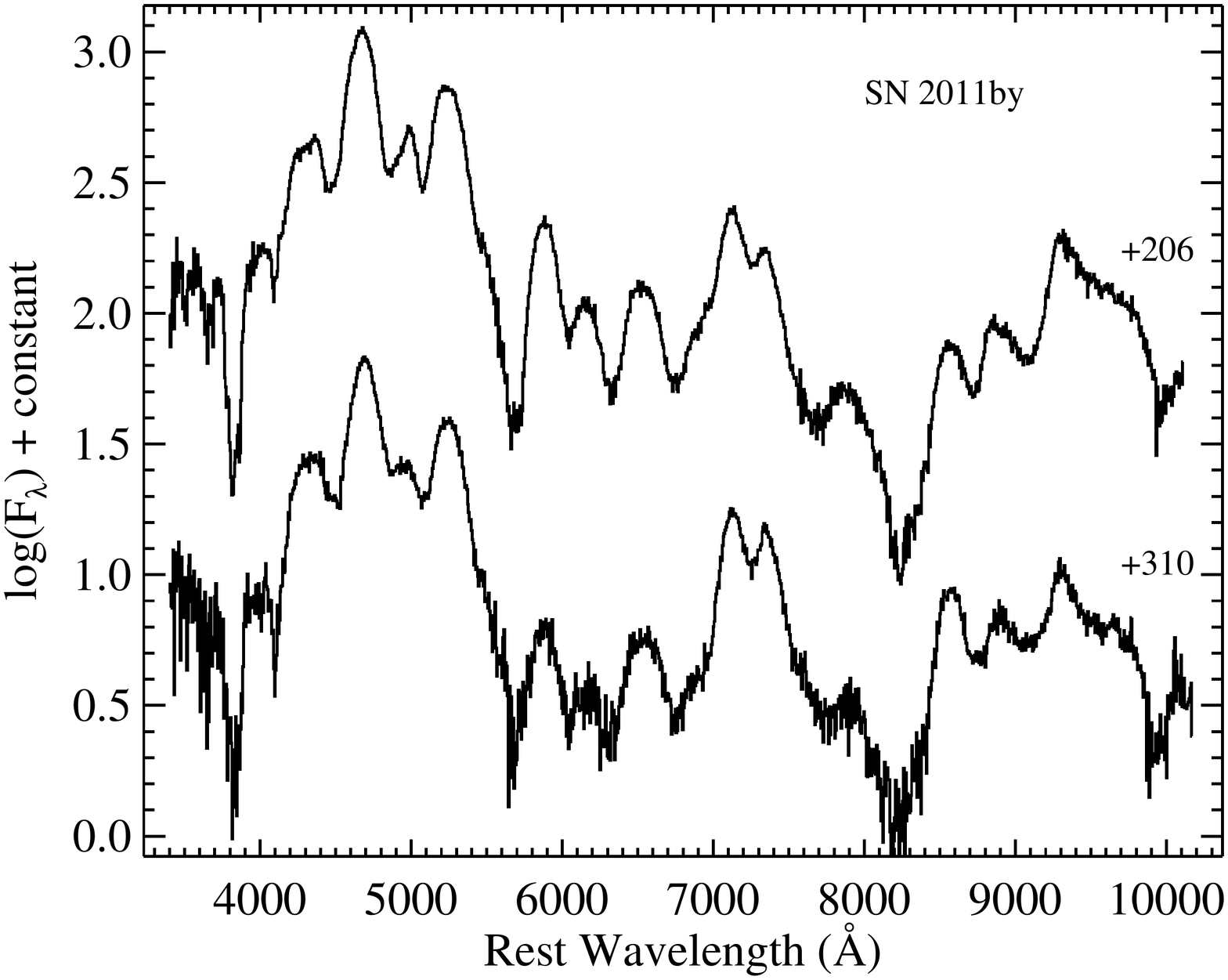} \\
\end{array}$
\caption{Near-maximum-brightness ({\it top}) and late-time ({\it
    bottom}) spectra of SN~2011by, labeled with age relative to
  the time of maximum brightness. The data have had their 
  host-galaxy recession velocity
  removed and have been corrected for Galactic
  reddening.}\label{f:sn2011by_spec}
\end{figure}

We obtained 7 near-maximum-brightness spectra of SN~2011by using
the Lick/Kast double spectrograph.
They span \about12~d before to \about15~d after maximum
brightness, as shown in the top panel of
Figure~\ref{f:sn2011by_spec}. In addition, we used Keck/LRIS 
to acquire 2
late-time spectra at 206 and 310~d past maximum (shown in the bottom
panel of Figure~\ref{f:sn2011by_spec}), the spectral feature
measurements of which can be found at the end of
Table~\ref{t:v_neb}. Details of all of our spectral observations of
SN~2011by can be found in Table~\ref{t:sn2011by}.

\setlength{\tabcolsep}{0.065in}
\begin{table}
\caption{Journal of spectroscopic observations of SN~2011by.\label{t:sn2011by}}
\begin{center}
\begin{tabular}{lrccc}
\hline\hline
UT Date &Age$^\textrm{a}$&Range (\AA)&Airmass$^\textrm{b}$&Exp (s)\\
\hline
2011 Apr.\ 28.2 & $-11.6$ & 3440--10200	 & 1.05	 & 600 \\
2011 May 4.2 & $-5.6$ & 3440--10200	 & 1.06	 & 600 \\
2011 May 6.2 & $-3.7$ & 3440--10200	 & 1.05	 & 600 \\
2011 May 10.2 & 0.3 & 3438--10196	 & 1.06	 & 600 \\
2011 May 14.2 & 4.3 & 3456--10200	 & 1.07	 & 600 \\
2011 May 22.3 & 12.3 & 3464--10200	 & 1.15	 & 600 \\
2011 May 25.3 & 15.4 & 3436--10200	 & 1.27	 & 600 \\
\hline
2011 Dec.\ 2.6$^\textrm{c}$ & 206 & 3400--10138	 & 1.50	 & 450 \\
2012 Mar.\ 15.5$^\textrm{c}$ & 310  & 3400--10200	 & 1.32	 & 900 \\
\hline\hline
\multicolumn{5}{p{3.1in}}{$^\textrm{a}$Rest-frame days relative to the date of $B$-band maximum brightness, 2011~May~9.9 (see Section~\ref{ss:sn2011by}).} \\
\multicolumn{5}{p{3.1in}}{$^\textrm{b}$Airmass at midpoint of exposure.} \\
\multicolumn{5}{p{3.1in}}{$^\textrm{c}$These observations used LRIS \citep{Oke95} on the 10-m Keck~I telescope. The others used the Kast spectrograph on the Lick 3-m Shane telescope \citep{Miller93}.} \\
\end{tabular}
\end{center}
\end{table}
\setlength{\tabcolsep}{6pt}

Consistent with the light curves, the spectra indicate that this object is quite
normal; the SuperNova IDentification code \citep[SNID;][]{Blondin07},
as implemented by \citet{Silverman12:BSNIPI}, was run on all of our
spectra of SN~2011by and it is definitively classified as 
spectroscopically normal. As in \citet{Silverman12:BSNIPII}, we
calculated the near-maximum-brightness velocity based on
the minimum of 
the \ion{Si}{II} $\lambda$6355 absorption feature and find that it 
shows a ``normal'' velocity near maximum brightness, as opposed to
some SNe~Ia which exhibit higher-than-average expansion velocities
\citep[e.g.,][]{Wang09}. Furthermore, since we have multiple spectra
near maximum brightness, we can calculate the velocity gradient
\citep[e.g.,][]{Benetti05}; SN~2011by falls squarely in the LVG
group.

Individual objects (some found by amateur astronomers and 
some by large-scale surveys) discovered in nearby galaxies soon after
explosion will be invaluable resources for extending our understanding
of SNe~Ia. SN~2011by represents an exquisite case study for the
current work.
In recent years there
have been a handful of other nearby SNe~Ia found soon after explosion
\citep[e.g.,][]{Nugent11,Foley12,Silverman12:12cg} that will also serve
as excellent individual case studies once late-time spectra have been
obtained and analysed.

\label{lastpage}

\end{document}